\documentclass[11pt]{article}
\usepackage{blois,epsfig}

\def\be{\begin{equation}}
\def\ee{\end{equation}}
\def\bea{\begin{eqnarray}}
\def\eea{\end{eqnarray}}

\begin{document}
\vspace*{4cm}
\title{THE DARK MATTER DISTRIBUTION IN THE CENTRAL REGIONS OF GALAXY CLUSTERS}

\author{D.J. SAND, T. TREU, G.P. SMITH, R.S. ELLIS}

\address{Department of Astronomy, Caltech,\\
Pasadena, CA  91125, USA}

\maketitle\abstracts{ Cosmological N-body simulations predict that
dark matter halos should have a universal shape characterized by a
steep, cuspy inner profile.  Here we report on a spectroscopic study
of six clusters each containing a dominant brightest cluster galaxy
(BCG) with nearby gravitational arcs.  Three clusters have both radial
and tangential gravitational arcs, whereas the other three display
only tangential arcs.  We analyze stellar velocity dispersion data for
the BCGs in conjunction with the arc redshifts and lens models to
constrain the dark and baryonic mass profiles jointly.  For those
clusters with radial gravitational arcs we were able to measure
precisely the inner slope of the dark matter halo and compare it with
that predicted from CDM simulations. }

\section{Introduction}

The Cold Dark Matter (CDM) paradigm for structure formation is
extremely successful in explaining observations of the universe on
large scales (e.g. Percival et al. 2001; Spergel et al. 2003; Croft et
al. 2002; Bahcall et al. 2003).  A vital tool within the CDM model is
that of N-body simulations which are able to infer the properties of
DM halos down to $\sim$kpc scales and which have predicted a
``universal'' shape for DM density profiles (at mass scales ranging
from dwarf galaxies to clusters of galaxies) that goes like
$\rho_{DM}\propto r^{-\beta}$ at small radii (e.g. Navarro, Frenk \&
White 1997; Moore et al. 1998; Power et al. 2003; Fukushige et
al. 2003).  Nearly all numerical work points to a value of $\beta$
between 1 and 1.5.

Observational verification of the DM density profile at various mass
scales is very important in confirming the CDM model.  At the galaxy
cluster scale, we have developed a technique to measure the DM density
profile by combining constraints from gravitational lensing and the
stellar velocity dispersion profile of a centrally located BCG.  This
allows us to disentangle luminous and dark components of the mass
distribution in the inner regions of galaxy clusters.

\begin{figure}
\begin{center}
\psfig{figure=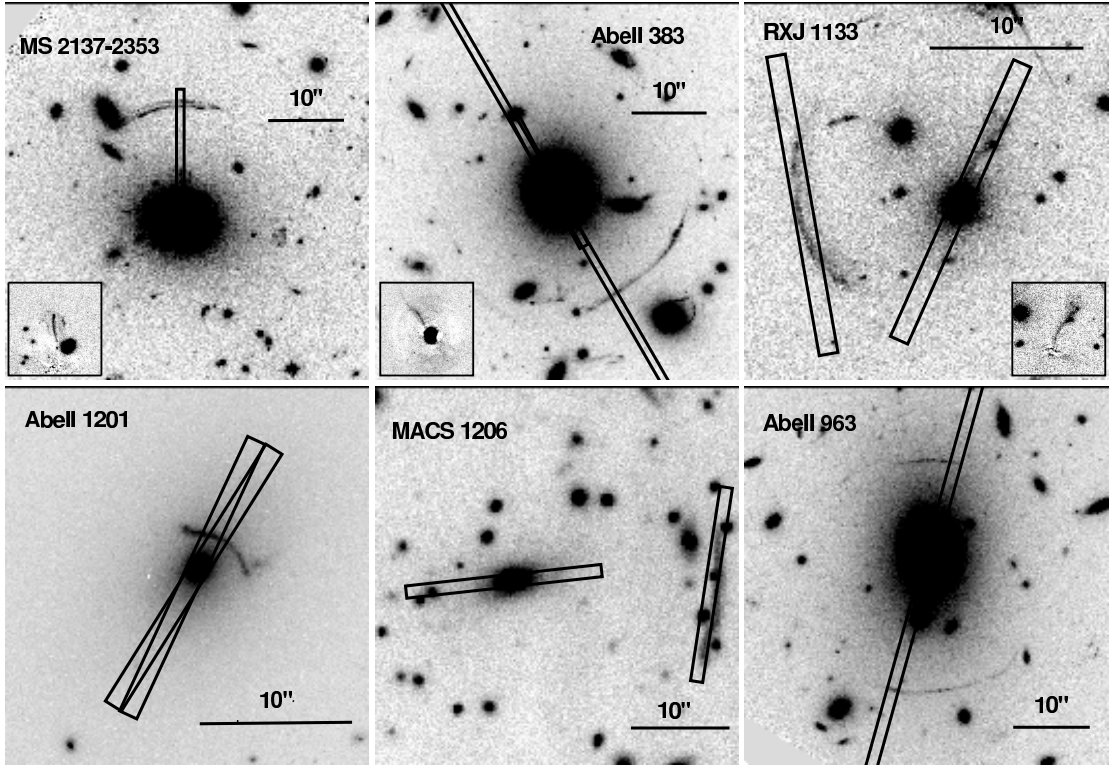,height=4.0in}
\end{center}
\caption{Images of the six clusters in this study.  The overlaid
``slits'' correspond to the actual slit positions and sizes that were
used.  The postage stamp insets show zoomed in, BCG-subtracted views
of the radial arcs.
\label{fig:radpdf}}
\end{figure}

\section{Sample Selection \& Data Analysis}

We have undertaken a search of the Hubble Space Telescope (HST) Wide
Field and Planetary Camera 2 (WFPC2) archive to look specifically for
radial gravitational arcs in galaxy clusters.  From this catalog of
clusters, we picked several candidates to follow up spectroscopically,
all of which met the following criteria: 1) the cluster had a
dominant, relatively isolated central galaxy; 2) there was at least
one nearby large tangential gravitational arc, although clusters with
radial arcs were preferred; 3) and there were no indications of
substantial substructure or a significantly elongated cluster
potential.  We collected Keck spectroscopic data for 6 galaxy
clusters, three with radial and tangential arcs and three with only
tangential arcs.  For each system an extended stellar velocity
dispersion profile of the BCG was measured and, if necessary,
gravitational arc redshifts were found as well.  Figure 1 shows an
image of each cluster.

We have presented the analysis involved in this work in Sand et
al. (2002, 2003); and only briefly describe the methodology here.  We
adopted a simple, spherically symmetric two-component mass model
comprising the BCG and cluster DM halo.  For the BCG component we used
a Jaffe (1983) mass density profile (which, in projection, describes
well the surface brightness profile of the BCGs) and for the cluster
DM halo we used a generalized-NFW profile which allows the inner
slope, $\beta$, to be a free parameter.  Given this mass model we have
effectively three free parameters: the stellar M/L of the BCG, the
amplitude of the DM density profile, $\delta_{c}$, and the crucial
inner DM density logarithmic slope, $\beta$.  For a given set of free
parameters, \{M/L,$\delta_{c}$,$\beta$\}, the expected position of the
gravitational arcs and stellar velocity dispersion of the BCG can be
calculated.  By comparing the expectations from a given mass model
with the observed arc positions and BCG velocity dispersion (along
with their uncertainties) a likelihood can be calculated.  In this
way, it is easy to marginalize with respect to various parameters to
find confidence limits on individual variables, in particular $\beta$.
Figure 2 gives the probability distribution function of the DM inner
density slope, $\beta$, for each of the clusters in our sample.  The
left hand panel shows the radial arc cluster constraints, along with
the joint distribution across all three systems
($\langle\beta\rangle=0.52^{+0.05}_{-0.05}$; 68\% CL).  Note not only
that the mean DM density profile is shallower than NFW, but that there
is significant scatter in $\beta$ values across the sample, $\Delta
\beta \sim$0.3.  The right hand panel shows the results for the
tangential arc sample ($\beta<$0.57; 99\% CL), which allows us to
place not only an upper limit on the value of $\beta$, but serves as a
useful control sample for the radial arcs systems.  Note that the
radial arc sample is not biased towards lower values of $\beta$ as
would be expected if they were a biased subset (see Sand et al. 2003).

\begin{figure}
\begin{center}
\psfig{figure=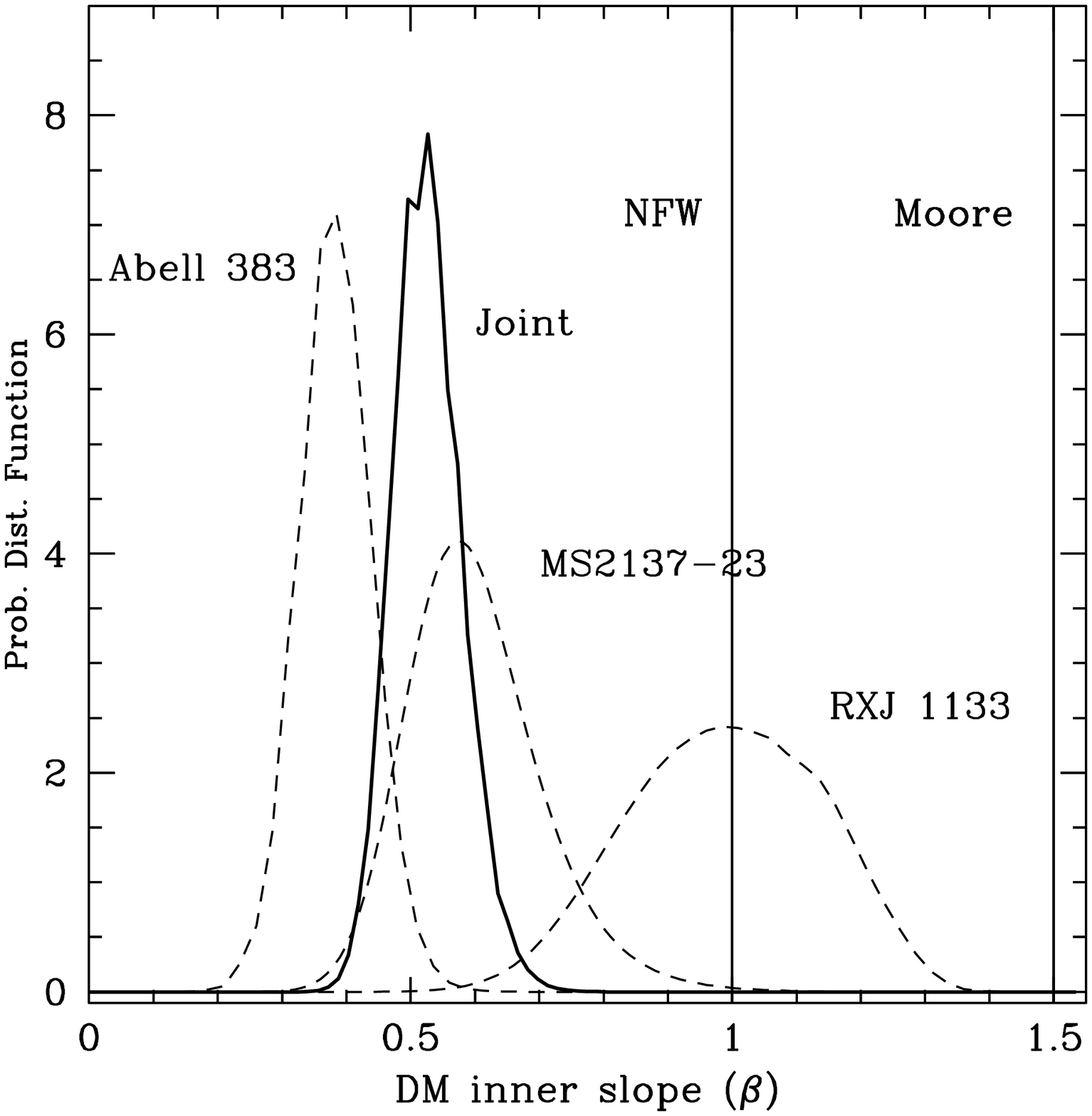,height=2.332in}
\psfig{figure=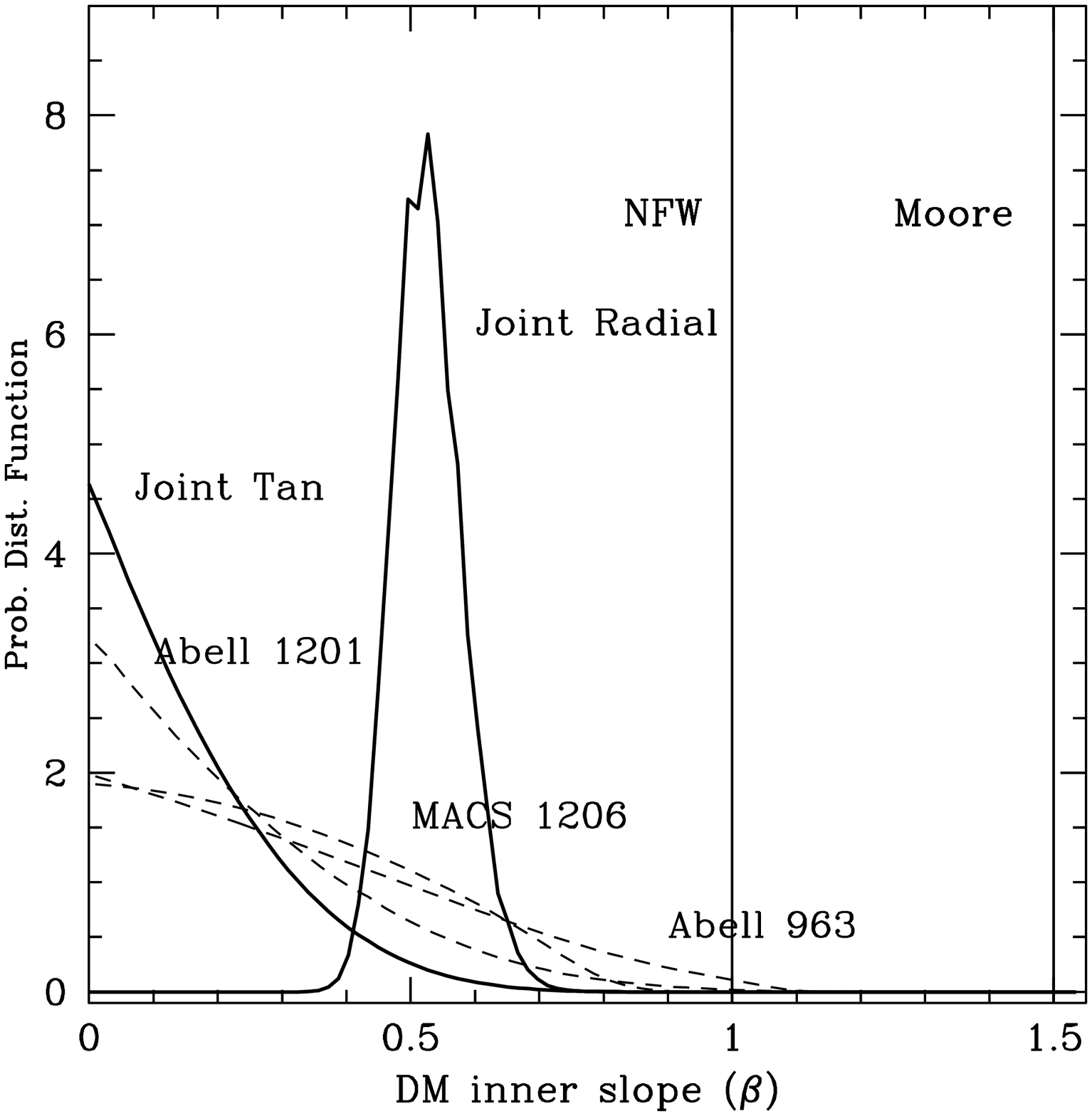,height=2.2in}
\caption{{\bf Left:} PDF of the DM inner density slope, $\beta$, for
the three radial arc clusters.  Note the wide scatter in $\beta$
values from cluster to cluster, $\Delta\beta\sim$0.3.  {\bf Right:}
PDF of the DM inner density slope, $\beta$, for the tangential arc
sample.  These allow us to place an upper limit on $\beta$ for each
cluster.
\label{fig:radpdf}}
\end{center}
\end{figure}

\section{Discussion \& Conclusions}

Figure 3 illustrates why we are able to place such strong constraints
on the shape of the DM profile.  In the left panel is plotted the
velocity dispersion profile measured for MS2137-23, along with the
best-fitting velocity dispersion (solid line; $\beta$=0.57) that
agrees with both the lensing and dynamics in the system.  Plotted as a
dashed line is a velocity dispersion profile whose underlying mass
distribution ($\beta$=1.30) agrees very well with the lensing data,
but does not fit the measured velocity dispersion.  This case
illustrates why mass models with too steep an inner profile cannot
match both the velocity dispersion profile and the positions of the
gravitational arcs.  In the right panel of Figure 3 is plotted the
best-fitting density profile of MS2137-23.  The velocity dispersion
measurement of the BCG allows us to probe the mass distribution where
luminous matter is important, while the arcs probe the portion of the
mass distribution where dark matter dominates.  The two measurements
complement each other.

The observed value of $\beta$ is expected to lie between 1 and 1.5 on
the basis of CDM only simulations, including those with the latest
refinements and consideration for numerical convergence (i.e. Power et
al. 2003; Fukushige et al. 2003).  We have found a range of acceptable
values of $\beta$ (see Figure 2), and although individual systems can
be consistent with NFW (e.g. RXJ 1133), the average slope is
inconsistent with the cuspy profiles expected from CDM simulations.
Is it possible to account for the discrepancy between these
observations and numerical predictions?  Conventional CDM simulations
only include collisionless DM particles.  It is not clear how the
inclusion of baryonic matter would affect the DM distribution,
especially in regions where it dominates the total matter density.
One possible situation, adiabatic contraction (e.g. Blumenthal et
al. 1986), would steepen the DM distribution through gravitational
processes and exacerbate the current problem.  Baryons could also play
an important dynamical role by driving energy and angular momentum out
of the cluster core, thus softening an originally cuspy profile
(e.g. El-Zant et al. 2001,2003).  It is also possible that the DM
particle is self-interacting.  This would naturally cause the DM
density profile to be shallower than that predicted from standard CDM
(Spergel \& Steinhardt 2000).

\begin{figure}
\begin{center}
\psfig{figure=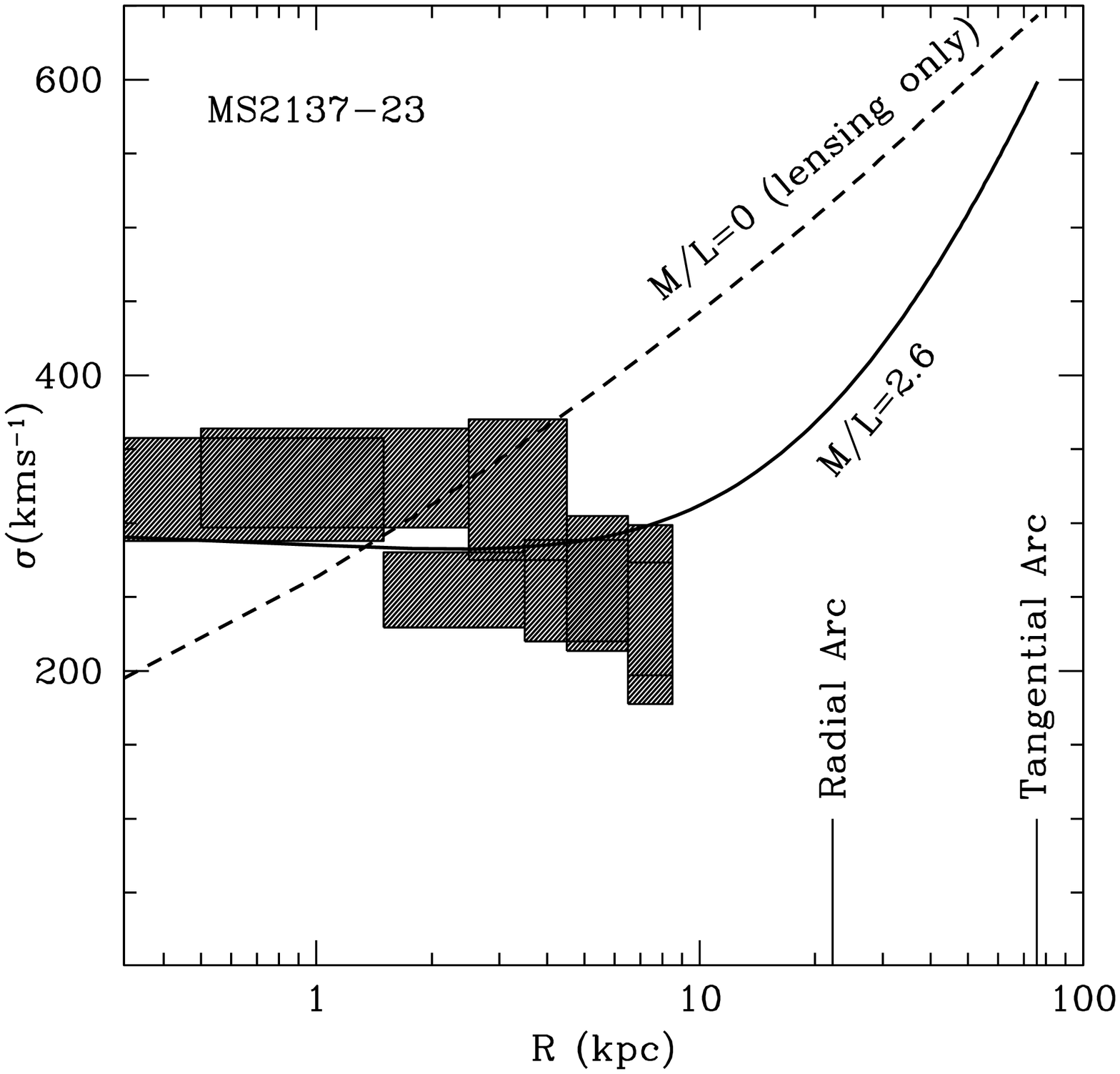,height=2.332in}
\psfig{figure=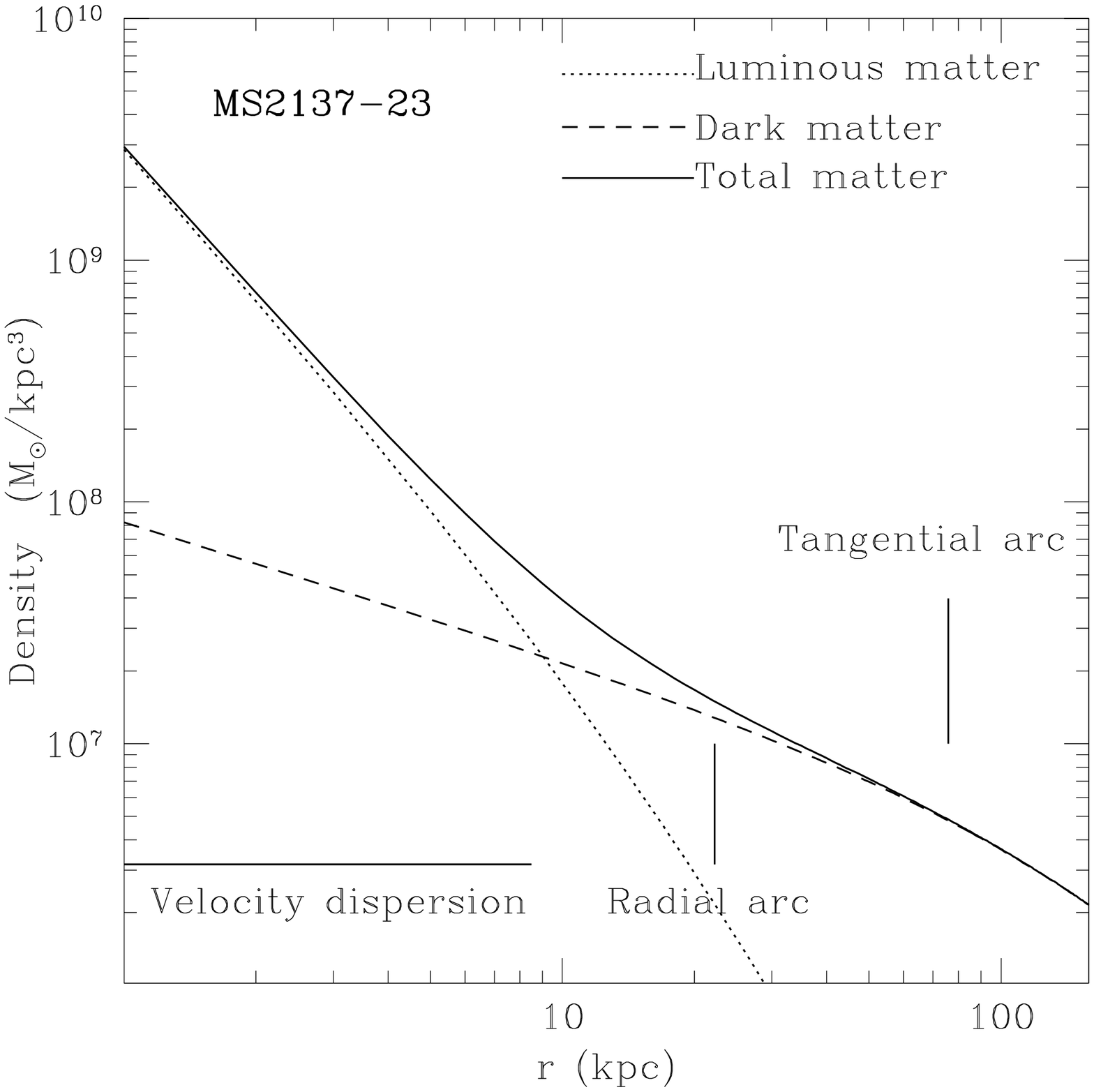,height=2.2in}
\caption{{\bf Left:} An illustration of why including the velocity
dispersion profile of BCG with the lensing analysis improves our
constraints on $\beta$.  The hatched boxes show the observed velocity
dispersion profile of MS2137-23.  The dashed curve shows a velocity
dispersion profile from a mass model that is compatible with the
lensing analysis, but not the dynamical measurement.  The solid curve
shows the best fitting profile obtained from the combined lensing +
dynamics analysis. {\bf Right:} Best-fitting total density profile for
the cluster MS2137-23.
\label{fig:MS2137}}
\end{center}
\end{figure}

We are in the process of collecting a larger sample of galaxy clusters
with radial arcs in order to further constrain our determination of
the mean value of $\beta$ and its intrinsic scatter.  A clear
measurement of both of these parameters will aid in future comparisons
to simulations, especially those that include both baryons and DM.

\section*{Acknowledgments}
\small
We are grateful to the MACS collaboration for access to MACS 1206.  We
acknowledge financial support for proposal number HST-AR-09527
provided by NASA through a grant from STScI, which is operated by
AURA, under NASA contract NAS5-26555.

\end{document}